\magnification \magstep1
\raggedbottom
\openup 4\jot
\voffset6truemm
\leftline {\bf COMPLEX PARAMETERS IN QUANTUM MECHANICS}
\vskip 1cm
\leftline {Giampiero Esposito}
\vskip 0.3cm
\noindent
{\it INFN, Sezione di Napoli, Mostra d'Oltremare Padiglione 20,
80125 Napoli, Italy}
\vskip 0.3cm
\noindent
{\it Universit\`a di Napoli Federico II, Dipartimento di Scienze
Fisiche, Complesso Universitario di Monte S. Angelo, Via Cintia,
Edificio G, 80126 Napoli, Italy}
\vskip 1cm
\noindent
The Schr\"{o}dinger equation for stationary states 
in a central potential is studied in
an arbitrary number of spatial dimensions, say $q$. After 
transformation into an equivalent equation, where the coefficient
of the first derivative vanishes, it is shown that in such equation
the coefficient of $r^{-2}$ is an even function of a parameter,
say $\lambda$, depending on a linear combination of $q$ and of the
angular momentum quantum number, say $l$. Thus, the case of complex
values of $\lambda$, which is useful in scattering theory, 
involves, in general, both a complex value of the parameter
originally viewed as the spatial dimension and complex
values of the angular momentum quantum number. The paper ends with
a proof of the Levinson theorem in an arbitrary number of spatial
dimensions, when the potential includes a non-local term which might
be useful to understand the interaction between two nucleons.
\vskip 100cm
\leftline {\bf 1. INTRODUCTION}
\vskip 0.3cm
Although the general framework of non-relativistic quantum 
mechanics is by now well known [1,2],
while its predictions have been
carefully tested against observations [3], it remains quite
important to understand whether some familiar problems are a
particular case of a more general scheme. For this purpose, we
here consider the Schr\"{o}dinger equation for stationary states
in an arbitrary number of spatial dimensions, say $q$. This topic
has always attracted interest [4--7],  
and we will see that a simple but deep result is found to hold
independently of the particular value of $q$. Section 2 is
devoted to a pedagogical formulation of the eigenvalue problem.
Section 3 derives in detail the reduction of the Schr\"{o}dinger
equation to a simpler form. General results on scattering states
are described in Sec. 4, and the Levinson theorem is proved in
Sec. 5 in an arbitrary number of dimensions and in the presence
of a non-local term in the potential.
\vskip 0.3cm
\leftline {\bf 2. STATIONARY STATES IN $q$ DIMENSIONS}
\vskip 0.3cm
As is well known, a potential on ${\cal R}^{q}$ which only depends
on the distance from the origin, say $r$, is called spherically
symmetric, or central. 
On denoting by $\bigtriangleup$ the Laplacian, we are
interested in its action on the domain $D$ of $C^{\infty}$ functions 
with compact support away from the origin: $D \equiv C_{0}^{\infty}
({\cal R}^{q}/ \{ 0 \})$. Each square-integrable function on
${\cal R}^{q}$ may be viewed as a function of $r$ and of $q-1$
variables $\xi$ on the $(q-1)$-sphere $S^{q-1}$. 
The solution by separation of variables of the
Schr\"{o}dinger equation for stationary states 
leads to the consideration of functions in $D$ which are
finite linear combinations of products $f(r)g(\xi)$. Such a set
is dense in the Hilbert space of square-integrable functions on
${\cal R}^{q}$, because [4]
$$
L^{2}({\cal R}^{q})=L^{2}({\cal R}_{+},r^{q-1}dr)
\otimes L^{2}(S^{q-1},d\Omega),
\eqno (2.1)
$$
where $d\Omega$ is the integration measure on $S^{q-1}$.
Equation (2.1) tells us that square-integrable functions on
${\cal R}^{q}$ belong to the tensor product of the space of
square-integrable functions on the positive half-line 
${\cal R}_{+}$ (the measure being $r^{q-1}dr$) with the space
of square-integrable functions on $S^{q-1}$.
On functions of the form $f(r)g(\xi)$, the operator
$-\bigtriangleup$ acts by [4]
$$
(-\bigtriangleup)f(r)g(\xi)=\left(-{d^{2}\over dr^{2}}
-{(q-1)\over r}{d\over dr}\right)f(r)g(\xi) 
-{1\over r^{2}}f(r)B_{q-1}g(\xi),
\eqno (2.2)
$$
where $B_{q-1}$ is the Laplace--Beltrami operator on 
$L^{2}(S^{q-1})$ [8]. The operator $B_{q-1}$ is found to be 
negative-definite, with only point spectrum of finite multiplicity,
and with $C^{\infty}$ eigenfunctions. The $l$-th eigenvalue,
$k_{l}$, is given by [4,8]
$$
k_{l}=-l(l+q-2).
\eqno (2.3)
$$
Thus, the Schr\"{o}dinger equation for stationary states [5]:
$$
\left(-{{\hbar}^{2}\over 2m} \bigtriangleup + U(r) \right)
\Psi({\vec r})=E \Psi({\vec r}),
\eqno (2.4)
$$
leads to the following equation for the radial part, $\psi(r)$,
of the wave function [5,7]: 
$$
\left[{d^{2}\over dr^{2}}+{(q-1)\over r}{d\over dr}
+{2m \over {\hbar}^{2}}(E-U(r))-{l(l+q-2)\over r^{2}}
\right]\psi(r)=0,
\eqno (2.5)
$$
on expressing the wave function as the product [5]
$$
\Psi({\vec r})=\psi(r)g(\theta_{1}, ..., \theta_{q-1}).
\eqno (2.6)
$$
The function $g$ in Eq. (2.6) is known as the generalized spherical
harmonic [5,8], and further indices have been omitted for
simplicity of notation. Of course, the full solution is eventually
obtained after summing all contributions of the kind (2.6),
the summation being taken over the whole set of quantum numbers
corresponding to the given value of $q$.
\vskip 0.3cm
\leftline {\bf 3. REDUCTION TO A SIMPLER FORM}
\vskip 0.3cm
One now wants to map Eq. (2.5) into an equation where the
coefficient of the first derivative vanishes, so that a problem
involving the Hilbert space of square-integrable functions on the
positive half-line ${\cal R}_{+}$ with respect to the measure
$r^{q-1}dr$ is mapped into a genuinely ``one-dimensional" problem,
in that one considers the square-integrable functions on the 
positive half-line ${\cal R}_{+}$ with respect to the measure $dr$.
For this purpose, one defines a new function $y$ by means of the
equation (see page 194 of Ref. [9])
$$
\psi(r) \equiv r^{\beta}y(r),
\eqno (3.1)
$$
where $\beta$ is, for the time being, an unknown parameter. The
request that, upon insertion of Eq. (3.1) into Eq. (2.5), one should
get a second-order differential equation for $y$ where the
coefficient of ${dy \over dr}$ vanishes, identifies $\beta$ as the
solution of the algebraic equation
$$
2\beta+(q-1)=0,
\eqno (3.2)
$$
and hence the desired transformation reads, eventually,
$$
\psi(r) \equiv r^{-{(q-1)\over 2}}y(r).
\eqno (3.3)
$$
On setting
$$
{\widetilde \kappa} \equiv {2mE \over {\hbar}^{2}},
\eqno (3.4)
$$ 
$$
V(r) \equiv {2m \over {\hbar}^{2}}U(r),
\eqno (3.5)
$$
the resulting differential equation for $y$ reads
$$
\left[{d^{2}\over dr^{2}}-\left(l(l+q-2)+{1\over 4}
(q^{2}-4q+3)\right){1\over r^{2}}+{\widetilde \kappa} \right]
y(r)=V(r)y(r).
\eqno (3.6)
$$
So far, what we have done is quite standard. However, unlike the
treatment in Refs. [5,6], where the coefficient of $r^{-2}$ is
expressed in a form proportional to $(p-1)(p-3)$, with
$p \equiv q+2l$, we would like to re-express it in a form which
is an even function of another suitable parameter, say $\lambda$,
because this is indeed the case when $q=3$, and one wonders what 
happens if the number of spatial dimensions is greater than 3.
For this purpose, we remark that the familiar technique of 
completing the square leads to
$$
l(l+q-2)=\left(l+{1\over 2}(q-2)\right)^{2}
-{1\over 4}(q-2)^{2},
\eqno (3.7)
$$
and hence one finds
$$
l(l+q-2)+{1\over 4}(q^{2}-4q+3)=
\left(l+{1\over 2}(q-2)\right)^{2}-{1\over 4}.
\eqno (3.8)
$$
Thus, the desired parameter, $\lambda$,
should be defined as
$$
\lambda \equiv l +{1\over 2}(q-2),
\eqno (3.9)
$$
so that Eq. (3.6) may be cast in the very convenient form
$$
\left[{d^{2}\over dr^{2}}+{\widetilde \kappa} 
-{(\lambda^{2}-{1\over 4})\over r^{2}}-V(r) \right]y(r)=0.
\eqno (3.10)
$$
It is now clear that, upon performing the transformation (3.3),
one obtains a second-order differential equation for $y$ in
{\it complete formal analogy} with what happens in 
${\cal R}^{3}$. In particular, for scattering states, 
one has $E>0$, and it is more convenient to define
$$
k^{2} \equiv {2mE \over {\hbar}^{2}},
\eqno (3.11)
$$
so that Eq. (3.10) reads
$$
\left[{d^{2}\over dr^{2}}+k^{2}
-{(\lambda^{2}-{1\over 4})\over r^{2}}-V(r)\right]y(r)=0.
\eqno (3.12)
$$
\vskip 0.3cm
\leftline {\bf 4. GENERAL RESULTS ON SCATTERING STATES}
\vskip 0.3cm
To appreciate the utility of Eq. (3.12), let us recall what one
learns from the analysis of the partial-wave boundary conditions 
at $r=0$ and at infinity [10]. The point $r=0$ is a Fuchsian
singularity [11] of Eq. (3.12), which admits two fundamental solutions,
say $\varphi_{1}$ and $\varphi_{2}$, such that [10]
$$
\varphi_{1}(r)=r^{\lambda + {1\over 2}} (1+{\rm o}(1))
\; \; {\rm as} \; \; r \rightarrow 0,
\eqno (4.1)
$$
$$
\varphi_{2}(r)=r^{-\lambda + {1\over 2}} (1+{\rm o}(1))
\; \; {\rm as} \; \; r \rightarrow 0.
\eqno (4.2)
$$
Since the operator in Eq. (3.12) is even in $\lambda$, the
fundamental solutions $\varphi_{1}$ and $\varphi_{2}$ exchange
their roles under the substitution $\lambda \rightarrow -\lambda$.
It is hence convenient to write $\varphi(\lambda,k,r)$ and 
$\varphi(-\lambda,k,r)$ in place of $\varphi_{1}(r)$ and
$\varphi_{2}(r)$, respectively, where the parameter $\lambda$ 
{\it is allowed to be freely specifiable in the complex plane} [10]. 

Moreover, as $r \rightarrow \infty$, one looks for the Jost
solution $f(\lambda,k,r)$ of Eq. (3.12), which satisfies
the asymptotic condition [10]
$$
\lim_{r \to \infty}e^{ikr}f(\lambda,k,r)=1.
\eqno (4.3)
$$
A deep relation exists between the boundary conditions
at $r=0$ and at $r=\infty$ [10]. In the former case, one finds
that $\varphi(\lambda,k,r)$ is entire (i.e. analytic in the 
whole complex plane) in $k^{2}$ and analytic in the domain
${\rm Re}(\lambda)>0$. If, for some positive $C$, 
$$
|V(r)| < C r^{-2+\varepsilon},
\eqno (4.4)
$$
the analyticity domain can be extended up to 
${\rm Re}(\lambda)> -{1\over 2}\varepsilon$. The Wronskian $W$
satisfies the property
$$
W \Bigr[\varphi(\lambda,k,r),\varphi(-\lambda,k,r)\Bigr]
=-2\lambda ,
\eqno (4.5)
$$
and, if $V$ is real-valued, the following Hermiticity condition
holds:
$$
\Bigr[\varphi(\lambda,k,r)\Bigr]^{*}
=\varphi(\lambda^{*},k^{*},r).
\eqno (4.6)
$$
In the latter case, one finds that $f(\lambda,k,r)$ is entire in
$\lambda^{2}$ and analytic in the domain ${\rm Im}(k)<0$. If the
potential satisfies the inequality
$$
|V(r)| < C e^{-u r},
\eqno (4.7)
$$
the analyticity domain of $f$ can be extended up to
${\rm Im}(k)<{u\over 2}$. The Wronskian yields
$$
W \Bigr[f(\lambda,k,r),f(\lambda,-k,r)\Bigr]=2ik,
\eqno (4.8)
$$
and, if $V$ is real-valued, the following Hermiticity 
condition holds:
$$
\Bigr[f(\lambda,k,r)\Bigr]^{*}
=f(\lambda^{*},-k^{*},r).
\eqno (4.9)
$$

In our $q$-dimensional problem, the form of Eq. (3.12) shows 
that all these properties still hold. However, by virtue of the
definition (3.9), it is $l+{q\over 2}-1$ which is allowed to
vary freely in the complex plane, and not merely $l$. For
example, from what we said above, 
the analyticity domain of $\varphi(\lambda,k,r)$ is
expressed by the condition
$$
{\rm Re} \left(l+{q\over 2}-1 \right)>0,
\eqno (4.10a)
$$
or, in equivalent form,
$$
{\rm Re}(l)+{1\over 2}{\rm Re}(q)>1.
\eqno (4.10.b)
$$
In other words, {\it the complex angular momentum formalism is actually
a particular case of a more general scheme}, where {\it both} $l$
{\it and} $q$ {\it are freely specifiable in the complex domain}. 
Interestingly, one can even keep $l$ real and still obtain complex
values of $\lambda$, provided that the parameter $q$ is allowed
to take complex values. Had one formulated from the beginning the
eigenvalue problem in $q$ dimensions, while bearing in mind the
properties of scattering theory in 3 dimensions, the above results
would have been clearer to the physics community.

Yet another interesting effect of the factor 
${1\over 2}(q-2)$ emerges in the analysis of potential scattering 
when the potential $V$ in Eq. (3.5) is assumed to admit a Laurent
expansion in the annulus $r \in ]0,\infty[$, so that Eqs. (2.5),
or (3.6) or (3.12) acquire non-Fuchsian singularities both as
$r \rightarrow 0$ and as $r \rightarrow \infty$ [7,12]. 
One then finds solutions of Eq. (2.5) in the form [7,12]
$$
r^{\gamma}\sum_{n=-\infty}^{\infty}c_{n}r^{n}.
$$
The fractional part of the 
parameter $\gamma$ can be found by solving a transcendental 
equation involving a Hill determinant which is an even periodic
function of [7]
$$
{\widetilde \gamma} \equiv \gamma +{1\over 2}(q-2).
\eqno (4.11)
$$
In other words, the polydromy parameter receives a
contribution from $q$, and one may well allow $q$ to take 
complex values at this stage. 
\vskip 0.3cm
\leftline {\bf 5. LEVINSON THEOREM WITH NON-LOCAL POTENTIALS}
\vskip 0.3cm
The ground is now ready for studying a non-trivial application
of the property expressed by our Eq. (3.12). Following Ref. [13],
we consider the stationary Schr\"{o}dinger equation in a central
potential when a non-local term contributes to such potential. This
is motivated by what one knows about the interaction between two
nucleons, which is purely local at sufficiently large distances,
but acquires a non-local nature as the two nucleons approach each
other. The novel feature in our analysis, with respect to Ref. [13],
is the arbitrary number of spatial dimensions, so that, for 
scattering states, we study the equation (cf. (3.12))
$$
\eqalignno{
\; & \left[{d^{2}\over dr^{2}}+k^{2}
-{(\lambda^{2}-{1\over 4})\over r^{2}}-V(r) \right]y(r) \cr
&=r^{{(q-1)\over 2}} \int U(r,r')y(r')r'^{{(q-1)\over 2}}dr'.
&(5.1)\cr}
$$
In Eq. (5.1), the kernel $U(r,r')$, also called, with little abuse
of notation, non-local potential, is taken to be real-valued,
continuous, symmetric, and vanishing at large distances [13,14]:
$$
U(r,r')=U(r',r),
\eqno (5.2a)
$$
$$
U(r,r') \sim {\rm O}(r^{-1}) \; \; {\rm as} \; \; 
r \rightarrow 0,
\eqno (5.2b)
$$
$$
U(r,r')=0 \; \; \forall r \geq r_{0},
\eqno (5.2c)
$$
where $r_{0}$ is a sufficiently large value of the radial coordinate
(as we said before). The local part, $V(r)$, of the potential, is
taken to be real-valued, continuous, and such that
$$
V(r) \sim {\rm O}(r^{-1}) \; \; {\rm as} \; \; 
r \rightarrow 0,
\eqno (5.3a)
$$
$$
V(r)=0 \; \; \forall r \geq r_{0}.
\eqno (5.3b)
$$
It is also useful to introduce a real parameter, say $\mu$, for which
one can write
$$
V(r,\mu)=\mu V(r), \; \; 
U(r,r',\mu)=\mu U(r,r').
\eqno (5.4)
$$
The idea is that, as $\mu$ ranges from $0$ through $1$, the rescaled
potentials $V(r,\mu)$ and $U(r,r',\mu)$ range from $0$ through the
original values $V(r)$ and $U(r,r')$. The radial equation (5.1) is
hence replaced by
$$
\eqalignno{
\; & \left[{\partial^{2}\over \partial r^{2}}+k^{2}
-{(\lambda^{2}-{1\over 4})\over r^{2}}-V(r,\mu)\right]
y_{k,\lambda}(r,\mu) \cr
&=r^{{(q-1)\over 2}}\int U(r,r',\mu) y_{k,\lambda}(r',\mu)
r'^{{(q-1)\over 2}}dr'.
&(5.5)\cr}
$$
We now consider Eq. (5.5) for two different values, say $k$
(with solution $y_{k,\lambda}$) and $\overline k$ (with solution
${\overline y}_{k,\lambda}$), and multiply the equations for
$y_{k,\lambda}(r,\mu)$ and ${\overline y}_{k,\lambda}(r,\mu)$ by
${\overline y}_{k,\lambda}(r,\mu)$ and $y_{k,\lambda}(r,\mu)$,
respectively. On taking the difference of the resulting equations,
one finds
$$
\eqalignno{
\; & {\partial \over \partial r}\left(y_{k,\lambda}
{\partial \over \partial r}{\overline y}_{k,\lambda}
-{\overline y}_{k,\lambda}{\partial \over \partial r}y_{k,\lambda}
\right)+\Bigr({\overline k}^{2}-k^{2}\Bigr)y_{k,\lambda} \;
{\overline y}_{k,\lambda} \cr
&=r^{{(q-1)\over 2}}y_{k,\lambda} \int U(r,r',\mu)
{\overline y}_{k,\lambda}(r',\mu)r'^{{(q-1)\over 2}}dr' \cr
&-r^{{(q-1)\over 2}}{\overline y}_{k,\lambda}
\int U(r,r',\mu)y_{k,\lambda}(r',\mu)
r'^{{(q-1)\over 2}}dr'.
&(5.6)\cr}
$$
Since, by regularity, both $y_{k,\lambda}$ and
${\overline y}_{k,\lambda}$ have vanishing limit as 
$r \rightarrow 0$, the integration of (5.6) over the interval
$[0,r_{0}]$ yields
$$
\eqalignno{
\; & \left[y_{k,\lambda}(r,\mu){\partial \over \partial r}
{\overline y}_{k,\lambda}(r,\mu)
-{\overline y}_{k,\lambda}(r,\mu){\partial \over \partial r}
y_{k,\lambda}(r,\mu)\right]_{r=r_{0}^{-}} \cr
&+\Bigr({\overline k}^{2}-k^{2}\Bigr)
\int_{0}^{r_{0}}y_{k,\lambda}(r',\mu)
{\overline y}_{k,\lambda}(r',\mu)dr' \cr
&=\int \int r^{{(q-1)\over 2}} r'^{{(q-1)\over 2}}
y_{k,\lambda}(r',\mu){\overline y}_{k,\lambda}(r,\mu)
\Bigr[U(r',r,\mu)-U(r,r',\mu)\Bigr]dr \; dr'=0,
&(5.7)\cr}
$$
where the right-hand side of (5.7) vanishes by virtue of the symmetry
condition (5.2a). Since ${\overline k} \not = k$ by hypothesis, we can
multiply both sides of (5.7) by
${1\over {y_{k,\lambda}(r_{0},\mu) 
{\overline y}_{k,\lambda}(r_{0},\mu)}}
{1\over ({\overline k}^{2}-k^{2})}$, which yields
$$
\eqalignno{
\; & {1\over ({\overline k}^{2}-k^{2})}
\left[{1\over {\overline y}_{k,\lambda}(r_{0},\mu)}
{\partial \over \partial r}
\left . {\overline y}_{k,\lambda}(r,\mu) \right |_{r_{0}}
-{1\over y_{k,\lambda}(r_{0},\mu)}{\partial \over \partial r}
\left . y_{k,\lambda}(r,\mu) \right |_{r_{0}} \right] \cr
&=-\int_{0}^{r_{0}}{y_{k,\lambda}(r',\mu)\over 
y_{k,\lambda}(r_{0},\mu)}
{{\overline y}_{k,\lambda}(r',\mu)\over 
{\overline y}_{k,\lambda}(r_{0},\mu)} dr'.
&(5.8)\cr}
$$
At this stage, on taking the limit of both sides of (5.8)
as ${\overline k} \rightarrow k$, one finds
$$
{\partial \over \partial E}\left[{1\over y_{E,\lambda}(r,\mu)}
{\partial \over \partial r}y_{E,\lambda}(r,\mu) 
\right]_{r=r_{0}^{-}}=-y_{E,\lambda}^{-2}(r_{0},\mu)
\int_{0}^{r_{0}}y_{E,\lambda}^{2}(r',\mu)dr' < 0,
\eqno (5.9)
$$
where the subscript $k$ for $y(r,\mu)$ has been replaced by
$E={{\hbar}^{2}\over 2m}k^{2}$, which is more convenient from now
on. Similarly, one finds
$$
{\partial \over \partial E}\left[{1\over y_{E,\lambda}(r,\mu)}
{\partial \over \partial r}y_{E,\lambda}(r,\mu) 
\right]_{r=r_{0}^{+}}=y_{E,\lambda}^{-2}(r_{0},\mu)
\int_{r_{0}}^{\infty}y_{E,\lambda}^{2}(r',\mu)dr' > 0.
\eqno (5.10)
$$
Equations (5.9) and (5.10) tell us that, as energy increases, the
logarithmic derivative of the radial function at $r_{0}^{-}$ 
{\it decreases monotonically}, whereas that at $r_{0}^{+}$ 
{\it increases monotonically}. This is an expression of the
Sturm--Liouville theorem in $q$ dimensions for non-local potentials
in a central field (cf. Ref. [13]). 

The matching condition at $r_{0}$ for the logarithmic derivative of
the radial function is, of course,
$$
A_{\lambda}(E,\mu) \equiv \left[{1\over y_{E,\lambda}(r,\mu)}
{\partial \over \partial r} y_{E,\lambda}(r,\mu)
\right]_{r=r_{0}^{-}}=\left[{1\over y_{E,\lambda}(r,\mu)}
{\partial \over \partial r}y_{E,\lambda}(r,\mu)\right]_{r=r_{0}^{+}}.
\eqno (5.11)
$$
The form (5.5) of the stationary Schr\"{o}dinger equation is indeed
quite involved, but some limiting cases are easily dealt with. For
example, for a free particle, which corresponds to a vanishing value
of $\mu$, one finds ($J_{\lambda}$ being the Bessel function of first
kind and order $\lambda$)
$$
y_{E,\lambda}(r,0)=\sqrt{{1\over 2}\pi kr} \;
J_{\lambda}(kr),
\eqno (5.12a)
$$
when $E >0$, and
$$
y_{E,\lambda}(r,0)=e^{-i \lambda \pi /2} 
\sqrt{{1\over 2}\pi \kappa r} \; J_{\lambda}(i \kappa r),
\eqno (5.12b)
$$
with $\kappa \equiv {\sqrt{-2mE}\over {\hbar}}$, if $E \leq 0$.

In the interval $[r_{0},\infty[$, both $V$ and $U$ vanish, and for
positive values of $E$ two oscillating solutions of (5.5) exist, so
that the general solution reads (cf. Ref. [13])
$$
y_{E,\lambda}(r,\mu)=\sqrt{{1\over 2}\pi k r}\Bigr[
J_{\lambda}(kr)\cos \eta_{\lambda}(k,\mu)
-N_{\lambda}(kr)\sin \eta_{\lambda}(k,\mu)\Bigr],
\eqno (5.13)
$$
where $\eta_{\lambda}(k,\mu)$ is the phase shift, and $N_{\lambda}$
is the Neumann function of order $\lambda$. The matching condition
(5.11) leads to a very useful formula for the phase shift, upon using
the result (5.13), i.e.
$$
\tan \eta_{\lambda}(k,\mu)={J_{\lambda}(kr_{0})\over 
N_{\lambda}(kr_{0})} 
{\left[A_{\lambda}(E,\mu)
-k {J_{\lambda}'(kr_{0})\over J_{\lambda}(kr_{0})}
-{1\over 2r_{0}}\right] \over
\left[A_{\lambda}(E,\mu)
-k {N_{\lambda}'(kr_{0})\over N_{\lambda}(kr_{0})}
-{1\over 2r_{0}}\right]}.
\eqno (5.14)
$$
Equation (5.14) provides the key tool for proving the Levinson 
theorem, jointly with a careful analysis of matching conditions.
Such a theorem leads to
a relation between the total number of bound states and
phase shifts at zero momentum. Here, following Ref. [13], we shall
agree that the phase shift is determined with respect to the phase
shift $\eta_{\lambda}(k,0)$ for a free particle, where, by
definition, one chooses $\eta_{\lambda}(k,0)=0$. With such a
convention, the phase shift is determined completely as $\mu$ 
increases from $0$ to $1$. 

If $E \leq 0$, the only square-integrable solution of (5.5) is
$$
y_{E,\lambda}=e^{i(\lambda+1)\pi /2}
\sqrt{{1\over 2}\pi \kappa r} \; 
H_{\lambda}^{(1)}(i \kappa r),
\eqno (5.15)
$$
where $H_{\lambda}^{(1)}$ is the standard notation for Hankel
functions of first kind and order $\lambda$. 
Thus, the right-hand side of the matching
condition (5.11) reads
$$
\left[{1\over y_{E,\lambda}(r,0)}{\partial \over \partial r}
y_{E,\lambda}(r,0)\right]_{r=r_{0}^{+}}
={i \kappa H_{\lambda}^{(1)}(i \kappa r_{0})\over
H_{\lambda}^{(1)}(i \kappa r_{0})}+{1\over 2r_{0}}.
\eqno (5.16)
$$
As $E \rightarrow 0^{-}$, the right-hand side of (5.16) reduces to
$$
\left(-\lambda+{1\over 2}\right){1\over r_{0}}
\equiv \rho_{\lambda},
\eqno (5.17)
$$
whereas it tends to $-\kappa$ as $E \rightarrow -\infty$. Moreover,
the solution (5.12b) satisfies the condition
$$
\left[{1\over y_{E,\lambda}(r,0)}{\partial \over \partial r}
y_{E,\lambda}(r,0) \right]_{r=r_{0}^{-}}
={i \kappa J_{\lambda}'(i \kappa r_{0})\over 
J_{\lambda}(i \kappa r_{0})}+{1\over 2r_{0}},
\eqno (5.18)
$$
whose right-hand side tends to
$$
\left(\lambda+{1\over 2}\right){1\over r_{0}}
\equiv {\widetilde \rho}_{\lambda} \; \; {\rm as}
\; \; E \rightarrow 0^{-}.
\eqno (5.19)
$$
Thus, no bound state exists when $\mu=0$. Unlike the two-dimensional
case studied in Ref. [13], not even the so-called ``half-bound states"
(for which $\rho_{\lambda}={\widetilde \rho}_{\lambda}$)
may occur, because $\lambda$ can never vanish if $l$ is $\geq 0$
and $q$ is bigger than $2$ (see (3.9)).

Another crucial remark is that, if $A_{\lambda}(0,\mu)$ decreases 
across the value $\rho_{\lambda}$ as $\mu$ increases, an overlap 
between two ranges of variation of the logarithmic derivative on the
two sides of $r_{0}$ occurs. Bearing in mind the Sturm--Liouville
theorem expressed by (5.9) and (5.10), such an overlap means that 
the matching condition (5.11) can only be satisfied by one particular 
value of the energy, and hence a scattering state is turned into a
bound state. In general, as $\mu$ increases from $0$ to $1$, 
each time $A_{\lambda}$ decreases across $\rho_{\lambda}$, a 
scattering state is turned into a bound state for the above reasons.
By contrast, each time $A_{\lambda}(0,\mu)$ increases across
$\rho_{\lambda}$, a bound state is turned into a scattering state.
The number of bound states is then equal to the number of times
that $A_{\lambda}(0)$ decreases across $\rho_{\lambda}$ as $\mu$
ranges from $0$ through $1$, minus the number of times that
$A_{\lambda}(0)$ increases across the value $\rho_{\lambda}$. The
next task is now to prove that this difference equals 
$\eta_{\lambda}(0)$, the phase shift at zero momentum, divided
by $\pi$. 

For this purpose, we have to evaluate $\tan \eta_{\lambda}(k,\mu)$
when $k << {1\over r_{0}}$, because
$$
\eta_{\lambda}(0,\mu)=\lim_{k \to 0} \eta_{\lambda}(k,\mu).
$$
By virtue of the exact result (5.14) and of the limiting behaviour
of Bessel functions at small argument, one finds, to lowest order
in $kr_{0}$,
$$
\eqalignno{
\; & \tan \eta_{\lambda}(k,\mu) \sim -{\pi (kr_{0})^{2\lambda}
\over 2^{2\lambda} \lambda \Gamma^{2}(\lambda)}
{\left[A_{\lambda}(0,\mu)-(\lambda+{1\over 2}){1\over r_{0}}\right]
\over \left[A_{\lambda}(0,\mu)
+(\lambda-{1\over 2}){1\over r_{0}}\right]} \cr
&= -{\pi (kr_{0})^{2\lambda}\over 2^{2\lambda} \lambda
\Gamma^{2}(\lambda)}
{\left[A_{\lambda}(0,\mu)+(\rho_{\lambda}-{1\over r_{0}}) 
\right] \over
\left[A_{\lambda}(0,\mu)-\rho_{\lambda}\right]},
&(5.20)\cr}
$$
where we have neglected, for simplicity, higher-order terms in
$kr_{0}$ in the denominator (strictly, higher-order terms should
be included in the small-$(kr_{0})$ expansions, whenever one has
exact cancellation of leading terms). Unlike Eq. (21) of Ref. [13],
our Eq. (5.20) expresses the tangent of the phase shift in terms
of a formula involving the $\Gamma$-function, because $\lambda$ is
not necessarily an integer. However, it remains true that
$\tan \eta_{\lambda}(k,\mu)$ tends to $0$ as $k \rightarrow 0$
(since $\lambda \geq 0$), and hence $\eta_{\lambda}(0,\mu)$ is
always equal to an integer multiple of $\pi$. This means that the
phase shift changes discontinuously. Besides, the exact formula
(5.14) can be used to prove that the phase shift increases 
monotonically as the logarithmic derivative $A_{\lambda}(E,\mu)$
decreases [13]. Thus, $\eta_{\lambda}(0,\mu)$ jumps by $\pi$ if,
for $k$ sufficiently small, $\tan \eta_{\lambda}(k,\mu)$ changes
sign as $A_{\lambda}(E,\mu)$ decreases. To sum up, when $\mu$
ranges from $0$ through $1$, whenever $A_{\lambda}(0,\mu)$ decreases
from near and larger than the value $\rho_{\lambda}$ to smaller
than $\rho_{\lambda}$, the denominator in (5.20) changes sign from
positive to negative, leading to a jump of $\eta_{\lambda}(0,\mu)$
equal to $\pi$. By contrast, whenever $A_{\lambda}(0,\mu)$ increases
across $\rho_{\lambda}$, $\eta_{\lambda}(0,\mu)$ jumps by $-\pi$. 
Bearing in mind what we said in the paragraph after Eq. (5.19), we
conclude that $\eta_{\lambda}(0)$ divided by $\pi$ is indeed 
equal to the number $n_{\lambda}$ of bound states [13]:
$$
\eta_{\lambda}(0)=n_{\lambda} \; \pi ,
\eqno (5.21)
$$
which is the form of the Levinson theorem with an arbitrary number
of spatial dimensions in the presence of non-local potentials 
(see (5.2{\it a})--(5.2{\it c})).

Needless to say, the analysis in the present section relies heavily 
on the work in Ref. [13], but provides a non-trivial application of
Eq. (3.12) and also a non-trivial generalization of the work in
Ref. [13]. There is therefore increasing 
evidence that the formulation of quantum-mechanical problems in
an arbitrary number of spatial dimensions may lead to new
developments in scattering theory with complex or real 
values of the parameter $\lambda$ defined in Eq. (3.9)
(whereas, in the past, more emphasis had been put
on large-$q$ expansions in quantum mechanics [5]).
\vskip 10cm
\leftline {\bf REFERENCES}
\vskip 0.3cm
\noindent
\item {1.}
P. A. M. Dirac, {\it The Principles of Quantum Mechanics}
(Clarendon Press, Oxford, 1958).
\item {2.}
C. J. Isham, {\it Lectures on Quantum Theory. Mathematical
and Structural Foundations} (Imperial College Press,
Singapore, 1995).
\item {3.}
M. Born, {\it Atomic Physics} (Blackie \& Son, London, 1969).
\item {4.}
M. Reed and B. Simon, {\it Methods of Modern Mathematical
Physics. II. Fourier Analysis and Self-Adjointness}
(Academic, New York, 1975).
\item {5.}
A. Chatterjee, {\it Phys. Rep.} {\bf 186}, 249 (1990).
\item {6.}
H. A. Mavromatis, {\it Am. J. Phys.} {\bf 66}, 335 (1998).
\item {7.}
G. Esposito, {\it J. Phys. A} {\bf 31}, 9493 (1998).
\item {8.}
A. Er\-d\'e\-lyi (E\-di\-tor), 
{\it Ba\-te\-man Ma\-nu\-scri\-pt Pro\-je\-ct, Hi\-gher
Tran\-scen\-den\-tal Func\-ti\-ons. II} (McGraw--Hill, New York, 1953).
\item {9.}
E. T. Whittaker and G. N. Watson, {\it Modern Analysis}
(Cambridge University Press, London, 1927).
\item {10.}
V. de Alfaro and T. Regge, {\it Potential Scattering}
(North Holland, Amsterdam, 1965).
\item {11.}
A. R. Forsyth, {\it Theory of Differential Equations. III}
(Dover, New York, 1959).
\item {12.}
S. Fubini and R. Stroffolini, {\it Nuovo Cimento} {\bf 37},
1812 (1965).
\item {13.}
S. H. Dong, X. W. Hou and Z. Q. Ma, {\it J. Phys. A} {\bf 31},
7501 (1998).
\item {14.} 
K. Chadan, {\it Nuovo Cimento} {\bf 10}, 892 (1958).
 
\bye